\documentclass[12pt]{iopart}

\usepackage{graphics}

\begin{document}

\title[TDI and Orbital Motion]{The Effects of Orbital Motion on LISA Time Delay Interferometry}

\author{Neil J. Cornish and Ronald W. Hellings}

\address{Department of Physics, Montana State University, Bozeman, MT 59717}

\begin{abstract}
In an effort to eliminate laser phase noise in laser
interferometer spaceborne gravitational wave detectors, several
combinations of signals have been found that allow the laser noise to be
canceled out while gravitational wave signals remain.  This process is
called {\it time delay interferometry} (TDI).  In the papers that defined
the TDI variables, their performance was evaluated in the limit that the
gravitational wave detector is fixed in space.   However, the performance
depends on certain symmetries in the armlengths that are available if the
detector is fixed in space, but that will be broken in the actual rotating
and flexing configuration produced by the LISA orbits. In this paper we
investigate the performance of these TDI variables for the real LISA
orbits.  First, addressing the effects of rotation, we verify Daniel
Shaddock's result that the {\it Sagnac} variables $\alpha(t)$, $\beta(t)$,
and $\gamma(t)$ will not cancel out the laser phase noise, and we also
find the same result for the symmetric Sagnac variable $\zeta(t)$.  The
loss of the latter variable would be particularly unfortunate since this
variable also cancels out gravitational wave signal, allowing instrument
noise in the detector to be isolated and measured. Fortunately, we have
found a set of more complicated TDI variables, which we call
$\Delta$-Sagnac variables, one of which accomplishes the same goal as
$\zeta(t)$ to good accuracy.  Finally, however, as we investigate the
effects of the flexing of the detector arms due to non-circular orbital
motion, we show that all variables, including the {\it interferometer}
variables, $X(t)$, $Y(t)$, and $Z(t)$, which survive the rotation-induced
loss of direction symmetry, will not completely cancel laser phase noise
when the armlengths are changing with time.  This unavoidable problem will
place a stringent requirement on laser stability of $\sim5{\rm Hz}/\sqrt{\rm Hz}$.
\end{abstract}

\section{Introduction}

Space gravitational wave detectors employing Michelson laser
interferometry between free-flying spacecraft differ in many ways from
their laboratory counterparts.  Among these differences is the fact that
the interferometer arms may not be maintained at equal lengths.  The laser
phase noise that cancels in laboratory detectors when the signals in the
two equal arms are subtracted from each other will not cancel when the
signals in the unequal arms of the spaceborne detectors are subtracted.
However, methods have been developed \cite{ghtf} that allow the various one-way
signals from the two arms to be combined to produce several variables that
are void of laser phase noise. The process of combining one-way signals
from the various arms of the constellation of three spacecraft has been
named {\it time delay interferometry}, or TDI. Previous work has
demonstrated how these variables perform in the limit that the arms of the
interferometer are fixed in space, but Daniel Shaddock~\cite{ds} has pointed
out that one set of variables, the so-called {\it Sagnac}
variables, will not cancel out laser phase noise when the detector is
rotating.  In this paper we will look at all TDI variables to see how they
perform when the armlengths are not direction-symmetric, as a result of
the rotation of the arms in the plane of the detector, and are not
time-translation symmetric, as a result of non-circular orbits for the end
masses of the detector.

\section{Canceling Laser Phase Noise in Static Detectors}

\begin{figure}[t]
\vspace{80mm}
\includegraphics{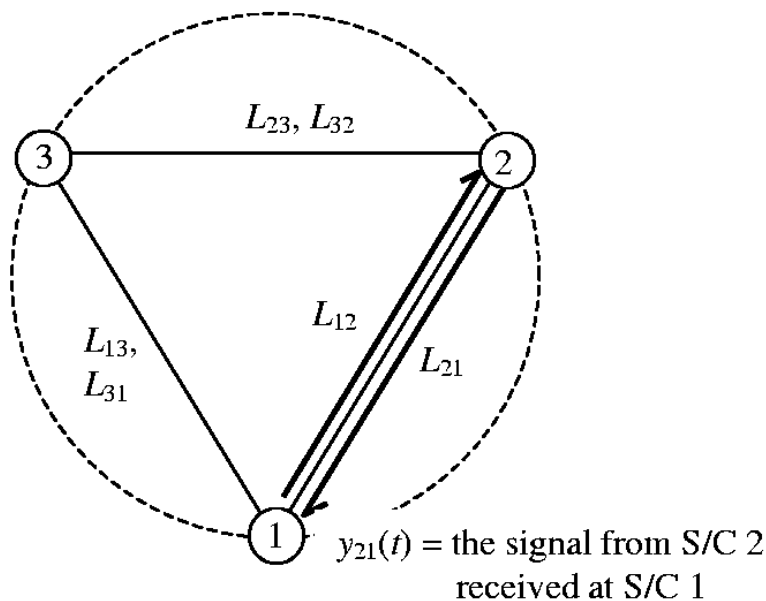}
\vspace{5mm}
\caption{The detector geometry and numbering convention for
spacecraft, armlengths, and signals.}
\end{figure}

Let us begin by showing how one may find variables that cancel laser phase
noise in the limit that the detector is static.  For reasons that will
soon be clear, we choose notation that follows that of Hellings \cite{ron}, as
shown in Figure 1. Three spacecraft, numbered 1, 2, and 3, are at the
vertices of a roughly equilateral triangle, with laser tracking signals
being continuously passed in both directions along each of the arms thus
formed. The one-way laser signal sent from spacecraft 2 and received at
spacecraft 1 at time $t$ is labeled $y_{21}(t)$.  The distance traveled
by the signal (in units of time) is labeled $L_{21}$. The signal passing
in the opposite direction along the same arm is labeled $y_{12}(t)$,
traveling a distance $L_{12}$.  Each laser signal $y_{ij}$ is formed by
beating the received laser light with light from the local laser, the
change in the phase of the beat signal being sensitive to the amplitude of
plane gravitational waves passing through the detector. If we write the
phase noise in the laser on spacecraft $i$ as $\phi_i(t)$, then the laser
phase noise in a one-way signal will be given by
\begin{equation}\label{phase}
y_{ij}=\phi_i(t-L_{ij})-\phi_j(t).
\end{equation}
Since the laser phase noise is many orders of magnitude larger than the
gravitational wave signal that is expected to be seen in the detector,
this noise must be eliminated.

There are actually several families of TDI variables that have been found
\cite{aet} that eliminate laser phase noise in a static detector.  The first one
we will consider is the family of {\it Sagnac variables} in which each
one-way signal appears once.  We write
\begin{eqnarray}\label{sagnac}
\sigma(t)&=&y_{12}(t-\lambda_1)+y_{23}(t-\lambda_2) +y_{31}(t-\lambda_3)\nonumber \\
&&-y_{13}(t-\lambda_4)-y_{32}(t-\lambda_5)-y_{21}(t-\lambda_6).
\end{eqnarray}
The reason for the name `Sagnac' for this variable is that it represents
the difference between the phase of a signal passed counter-clockwise
around the triangle, starting with spacecraft 1 (the first line of Eq.~\ref{sagnac}),
and one passed clockwise around the triangle (the second line),
starting at the same point. The time delays $\lambda_i$ are to be adjusted
to eliminate laser phase noise.  To see how this is done, let each term in
Eq.~\ref{sagnac} be expanded using Eq.~\ref{phase}, one line per term, to give
\begin{eqnarray}\label{sigmae}
\sigma(t)&=&\phi_1(t-\lambda_1-L_{12})-\phi_2(t-\lambda_1)\nonumber \\
 &+&\phi_2(t-\lambda_2-L_{23})-\phi_3(t-\lambda_2)\nonumber \\
 &+&\phi_3(t-\lambda_3-L_{31})-\phi_1(t-\lambda_3) \nonumber \\
 &-&\phi_1(t-\lambda_4-L_{13})+\phi_3(t-\lambda_4) \\
 &-&\phi_3(t-\lambda_5-L_{32})+\phi_2(t-\lambda_5)\nonumber \\
 &-&\phi_2(t-\lambda_6-L_{21})+\phi_1(t-\lambda_6)  .\nonumber
\end{eqnarray}
If the laser phase noise is to cancel exactly, we must have
\numparts
\begin{eqnarray}
\fl \quad \quad &&\phi_1(t-\lambda_1-L_{12})-\phi_1(t-\lambda_3)
-\phi_1(t-\lambda_4-L_{13})+\phi_1(t-\lambda_6)=0 \label{eg} \\
\fl \quad \quad -&&\phi_2(t-\lambda_1)+\phi_2(t-\lambda_2-L_{23})
+\phi_2(t-\lambda_5)-\phi_2(t-\lambda_6-L_{21})=0  \\
\fl \quad \quad -&&\phi_3(t-\lambda_2)+\phi_3(t-\lambda_3-L_{31})
+\phi_3(t-\lambda_4)-\phi_3(t-\lambda_5-L_{32})=0 
\end{eqnarray}
\endnumparts
There are two ways in which the four terms in Eq.~\ref{eg}, for example, may be
made to cancel.  One must either have $\lambda_1-L_{12}=\lambda_3$ and
$\lambda_4-L_{13}=\lambda_6$ or $\lambda_1-L_{12}=\lambda_4-L_{13}$ and $\lambda_3=\lambda_6$. There are
thus two equations for the $\lambda_i$ arising out of Eq.~\ref{eg}, and there are
two possibilities for the form of these two equations.  From Eqs.~4,
therefore, we will have eight ($2\times2\times2$) possible sets of six
equations for the six unknowns $\lambda_i$.  The six equations, however, will
always possess one degeneracy in them, corresponding to the freedom in the
origin of $t$.  Thus, Sagnac variables will be found as solutions to any
of the eight possible sets of six equations for five unknowns.

When the possible ways of satisfying Eqs.~4 are studied, with the
simplification $L_{ij}=L_{ji}$ that is appropriate to a static detector,
there appear only four solutions.  One of these is found by setting $\lambda_6$
to zero and solving for the remaining $\lambda_i$, giving
\begin{eqnarray}\label{sol1}
&&\lambda_1=L_{13}+L_{23} \nonumber \\
&&\lambda_2=L_{13}\nonumber \\
 &&\lambda_3=0 \\
 &&\lambda_4=L_{12}+L_{23}\nonumber \\
 &&\lambda_5=L_{12}\nonumber \\
&&\lambda_6=0 \nonumber 
\end{eqnarray}
corresponding to TDI variable $\alpha(t)$ of Armstrong, Estabrook, and
Tinto \cite{aet}.  Two other solutions may be found from Eq.~\ref{sol1} by simultaneously
permuting $1\rightarrow2\rightarrow3$ and $4\rightarrow5\rightarrow6$ on
the left hand side and $1\rightarrow2\rightarrow3$ on the right.  A single
permutation gives TDI variable $\beta(t)$ and a further permutation gives
$\gamma(t)$. The final possible solution to Eqs.~4, arbitrarily choosing
$\lambda_6=L_{23}$, is
\begin{eqnarray}
&&\lambda_1 = L_{13} \nonumber \\
&&\lambda_2=L_{12}\nonumber \\
&&\lambda_3=L_{23} \\
&&\lambda_4=L_{12}\nonumber \\
&&\lambda_5=L_{13}\nonumber \\
&&\lambda_6=L_{23}\nonumber
\end{eqnarray}
corresponding to the so-called {\it symmetric Sagnac} variable $\zeta(t)$.
The reason for the existence of solutions to six equations for five
unknowns is that symmetries arise in the equations due to the fact that
the direction symmetry $L_{ij}=L_{ji}$ reduces from six to three the
number of independent constants in the six equations arising out of Eqs.~4.

A second class of TDI variables consists of the so-called {\it
interferometer variables}, formed by choosing two arms out of the three
and using the four one-way signals going up and down these arms, with each
signal appearing twice.  For example, if we choose the two arms radiating
from spacecraft 1, we would form
\begin{eqnarray}\label{X}
X(t)&= &y_{12}(t-\lambda_1)-y_{13}(t-\lambda_2)+y_{21}(t-\lambda_3)
-y_{31}(t-\lambda_4)\nonumber \\
&&+y_{13}(t-\lambda_5)-y_{12}(t-\lambda_6)+y_{31}(t-\lambda_7)-y_{21}(t-\lambda_8)
\end{eqnarray}
When Eq.~\ref{X} is expanded using Eq.~\ref{phase}, one line per term, we write the laser
phase noise in $X(t)$ as
\begin{eqnarray}\label{Xe}
X(t)& =&\phi_1(t-\lambda_1-L_{12})-\phi_2(t-\lambda_1)\nonumber \\
 &&-\phi_1(t-\lambda_2-L_{13})+\phi_3(t-\lambda_2)\nonumber \\
 &&+\phi_2(t-\lambda_3-L_{21})-\phi_1(t-\lambda_3)\nonumber \\
 &&-\phi_3(t-\lambda_4-L_{31})+\phi_1(t-\lambda_4)\nonumber \\
 &&+\phi_1(t-\lambda_5-L_{13})-\phi_3(t-\lambda_5) \\
 &&-\phi_1(t-\lambda_6-L_{12})+\phi_2(t-\lambda_6)\nonumber \\
 &&+\phi_3(t-\lambda_7-L_{31})-\phi_1(t-\lambda_7)\nonumber \\
 &&-\phi_2(t-\lambda_8-L_{21})+\phi_1(t-\lambda_8)  .\nonumber
\end{eqnarray}
As in Eqs.~4, we again have three conditions for the identical vanishing
of the three laser phase noises, but the counterpart of Eq.~\ref{eg}, the
$\phi_1$ equation, contains the eight occurrences of $\phi_1$ that appear
in Eq.~\ref{Xe}.  The cancellation by pairs in this equation may occur 24
different ways, so there will be 96 ($24\times2\times2$) different sets of
eight equations for the seven unknown $\lambda_i$.  Whether
$L_{ij}=L_{ji}$ or $L_{ij}\neq L_{ji}$, the only unique, nontrivial solution to
these sets of equations is
\begin{eqnarray}\label{Xsol}
&&\lambda_1=L_{13}+L_{31}+L_{21}\nonumber \\
&& \lambda_2=L_{12}+L_{21}+L_{31}\nonumber \\
&&\lambda_3=L_{31}+L_{13}\nonumber \\
&&\lambda_4=L_{21}+L_{12}\nonumber \\
&& \lambda_5=L_{31} \\
&&\lambda_6=L_{21}\nonumber \\
&&\lambda_7=0\nonumber \\
&&\lambda_8=0 ,\nonumber
\end{eqnarray}
corresponding to the $X(t)$ interferometer variable found by Armstrong,
Estabrook, and Tinto~\cite{aet}.  If the two arms are chosen differently, so that
Eq.~\ref{Xe} is modified by the permutations $1\rightarrow2\rightarrow3$, then
the single solution each time will correspond to $Y(t)$ and then to
$Z(t)$.

Besides the {\it interferometer} variables of Eq.~\ref{X}, there are other ways
to combine the six possible one-way links into eight-term combinations.
These have been named the {\it beacon}, {\it monitor}, and {\it relay}
variables and are described in the appendix of Estabrook, Tinto, and
Armstrong~\cite{eta}.

\section{The Effects of Rigid Rotation}

All of the variables discussed above have the property that laser phase
noise exactly cancels out when the armlengths are direction-symmetric,
that is, when $L_{ij}=L_{ji}$.  When these variables were derived, this
direction symmetry was always assumed.  Indeed, the notation used in the
defining papers \cite{aet} gave a single subscript to each armlength (so, for
example, $L_1$ was the length of the arm {\it across} from spacecraft 1,
the length we have called $L_{23}$). However, there is a problem with the
assumption of direction symmetry.  This is that the constellation for the
LISA mission \cite{dr} will be rotating once per year in the plane of the
detector, due to the individual orbits of each end spacecraft. Since this
is the case, it will not be true that $L_{ij}=L_{ji}$, as we shall now
show.

\begin{figure}[t]
\vspace{65mm}
\includegraphics{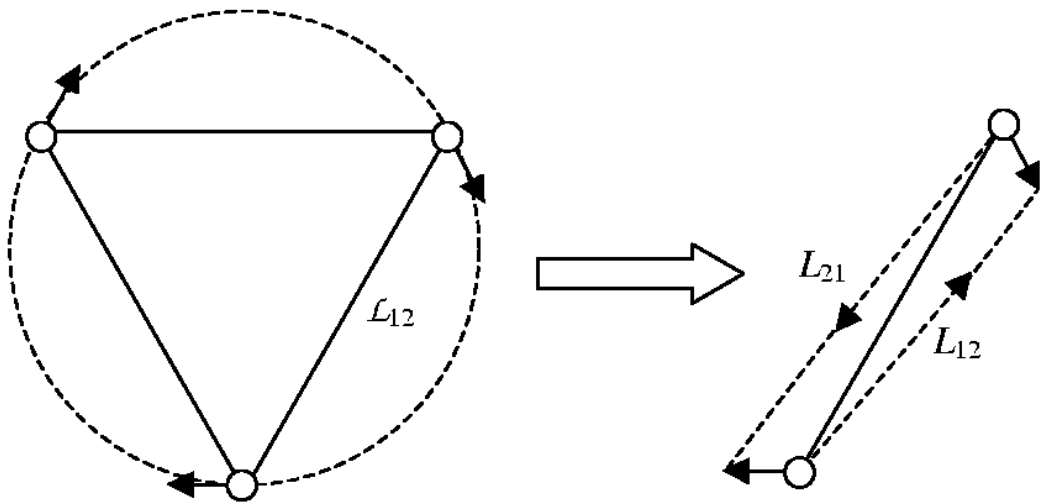}
\vspace{5mm}
\caption{How rotation of the detector breaks the direction symmetry
in the armlengths. See text for discussion.}
\end{figure}

Consider spacecraft 1 and 2 in Figure 2.  Since the constellation is
rotating in the clockwise direction, the signal sent from spacecraft 1 to
spacecraft 2 at time $t$ will have to aim ahead of the position of
spacecraft 2 at time $t$.  If we define the length of the arm between
spacecraft 1 and 2 to be ${\cal L}_{12}$ in the limit of no rotation, then the
actual distance traveled by the signal from spacecraft 1 to spacecraft 2
will be $L_{12}<{\cal L}_{12}$.  Similarly, the signal from spacecraft 2 to
spacecraft 1 will have to lead spacecraft 1 in its motion and will
therefore travel a distance $L_{21}>{\cal L}_{12}$.  When the constellation is
rotating, we will therefore not have $L_{ij}=L_{ji}$, and so the
laser-canceling variables will not work.  Indeed, since these variables
were all derived with the assumption $L_{ij}=L_{ji}$, it is not even clear
whether $L_{ij}$ or $L_{ji}$ should be used in the definitions.

In what follows, we will investigate the families of laser-canceling
variables for the case of direction asymmetry ($L_{ij}\neq L_{ji}$), but
in the limit of rigid rotation where the armlengths are constant in time.
If exact cancellation is not possible, we will estimate the size of the
residual laser phase noise and calculate the requirement for laser
frequency stability that this uncanceled noise will impose on the LISA
laser stabilization system.

\subsection{Sagnac-type Variables}

When $L_{ij}\neq L_{ji}$, none of the eight possible sets of six equations
for five unknowns represented in Eqs.~4 have solutions.  There is no
Sagnac-type or symmetric-Sagnac-type variable that will exactly cancel
laser phase noise. To see how much noise is left over, let us set the
$\lambda_i$ in Eqs.~4 so as to cancel as many terms as possible and then
evaluate the power spectral density of the noise that will remain.  We
begin with an $\alpha$-like variable by arbitrarily setting $\lambda_6=0$. Then
$\lambda_3=0$ will allow the second and last terms in Eq.~\ref{eg} to cancel.
Setting $\lambda_5=L_{21}$ will then allow the last two terms in Eq.~4b to
cancel, and so on. Working this way, we are able to cancel all terms but
two. We are left with
\begin{equation}\label{alpha}
\alpha(t)=\phi_1(t-L_{12}-L_{23}-L_{31})-\phi_1(t-L_{13}-L_{32}-L_{21})
\end{equation}
When $L_{ij}\neq L_{ji}$, these two terms do not cancel, and there remains
an unavoidable laser phase noise in the Sagnac variable $\alpha(t)$.
Similar results apply for $\beta(t)$ and $\gamma(t)$.

How big is the error produced by these two terms?  First, let us note that
$L_{12}+L_{23}+L_{31}$ is the total time around the constellation in the
counter-clockwise direction and that $L_{13}+L_{32}+L_{21}$ is the total
time in the clockwise direction.  Even if the triangle were perfectly rigid,
the times of flight would not be the same because of
the rotation.  Let us write the time of flight around the
triangle in the limit of no motion as ${\cal L}_{\rm tot}$, and the times of
flight in the counter-clockwise and clockwise directions as ${\cal L}_{\rm
tot}+\Delta L_-$ and ${\cal L}_{\rm tot}+\Delta L_+$, respectively, when the
triangle is rotating.  Since the LISA constellation rotates in a clockwise
direction (seen from the ecliptic pole), $\Delta L_-$ will always be
negative and $\Delta L_+$ will always be positive.  If we now expand the
phase noise in Eq.~\ref{alpha} in a Taylor series, we find
\begin{eqnarray}\label{alphae}
\alpha(t)&=&
\phi_1(t-{\cal L}_{\rm tot}-\Delta L_-)
   -\phi_1(t-{\cal L}_{\rm tot}-\Delta L_+) \nonumber \\
  & =&\phi_1(t-{\cal L}_{\rm tot})-\dot\phi_1\Delta L_-
   -\phi_1(t-{\cal L}_{\rm tot})+\dot\phi_1\Delta L_+  \\
  & =&(\Delta L_+-\Delta L_-)\dot\phi_1 \nonumber
\end{eqnarray}
The difference $\Delta L_+-\Delta L_-$ is the Sagnac time shift for
signals circulating around a closed path and is given by
\begin{equation}
\Delta L_+-\Delta L_-=\Delta L_{\rm Sagnac}={4A\Omega\over c2}
 ={\sqrt{3}L22\pi\over c2 T} 
\end{equation}
where $A$ is the area enclosed by the light path, $\Omega$ is the angular
velocity of the rotating light path, $L$ is a typical armlength, and $T$
is the period of the rotation of the detector.  For the LISA mission
parameters, this time difference is $\Delta L_{\rm Sagnac}=10^{-4}\,$s.
The relationship between residual noise power in the $\alpha(t)$ variable
and the laser frequency noise power is found from Eq.~\ref{alphae} to be
\begin{equation}
S_\alpha=S_\nu(\Delta L_{\rm Sagnac})2 
\end{equation}
so, for a requirement of $\sqrt{S_\alpha}= 10^{-5}\, {\rm
cycles}/\sqrt{{\rm Hz}}$, we derive a laser stability requirement of
\begin{equation}\label{sagreq}
\sqrt{S_\nu}={\sqrt{S_\alpha}\over\Delta L_{\rm Sagnac}}
={10^{-5}{\rm cycles}/\sqrt{{\rm Hz}}\over10^{-4}\, {\rm s}}=0.1\, {\rm
Hz}/\sqrt{{\rm Hz}} . 
\end{equation}
This is a very difficult requirement to satisfy.

Beginning again with Eq.~\ref{sigmae}, let us try to find a $\zeta$-like variable
that cancels as much laser phase noise as possible. Setting $\lambda_6=L_{31}$
and $\lambda_5=L_{32}$, we cancel the first terms in the last two lines of Eq.~\ref{sigmae}
Continuing in this way, we may choose the remaining $\lambda_i$ so as to
cancel all $\phi_2$ and $\phi_3$ terms.  We are left with
\begin{equation} \label{zeta}
\fl \quad \quad \zeta(t)=\phi_1(t-L_{31}-L_{12})
-\phi_1(t-L_{21}-L_{13})+\phi_1(t-L_{23})-\phi_1(t-L_{32}) .
\end{equation}
If we write each armlength as $L_{ij}={\cal L}_{ij}+\Delta L_{ij}$, where
${\cal L}_{ij}={\cal L}_{ji}$ is the armlength in a co-rotating system, and expand
$\phi_1$ in a Taylor series, Eq.\ref{zeta} becomes
\begin{eqnarray}\label{zetae}
\fl \quad \quad \zeta(t)&=&\phi_1(t-{\cal L}_{31}-{\cal L}_{12})-(\Delta L_{31}+\Delta L_{12})\dot\phi_1
 -\phi_1(t-{\cal L}_{21}-{\cal L}_{13})+\nonumber \\
\fl &&(\Delta L_{21}+\Delta L_{13})\dot\phi_1
 +\phi_1(t-{\cal L}_{23})-\Delta L_{23}\dot\phi_1-\phi_1(t-{\cal L}_{32})+\Delta L_{32}\dot\phi_1 .
\end{eqnarray}
Collecting terms, we have
\begin{equation}
\zeta(t)=\left[(\Delta L_{13}+\Delta L_{32}+\Delta L_{21})
-(\Delta L_{12}+\Delta L_{23}+\Delta L_{31})\right]\dot\phi_1
\end{equation}
This is the same Sagnac path difference as for $\alpha(t)$, so the
uncanceled laser phase noise in $\zeta(t)$ will be the same as that
derived for $\alpha(t)$, $\beta(t)$, and $\gamma(t)$, with the same
resulting requirement for laser frequency stability that we derived in
Eq.~\ref{sagreq}.  Since this requirement is not likely to be satisfied, we must
draw the conclusion that the $\zeta(t)$ variable will simply not be
available.  This would be particularly unfortunate, since $\zeta(t)$ has
the interesting property that low-frequency-limit gravitational-wave
signals exactly cancel out when this combination is formed \cite{aet} so that it
would have provided the potentially useful capability of differentiating
between the power produced by instrument noise, which would still remain
in $\zeta(t)$, and that produced by a gravitational wave background, which
would go to zero in $\zeta(t)$ but not in the other variables.  The
variable $\zeta(t)$ will not accomplish this goal for a rotating detector,
but we have found a more complicated variable that will accomplish the
same goal to good accuracy.  This $\Delta\zeta(t)$ variable is discussed
in the next section.

\subsection{$\Delta$-Sagnac-type Variables}

The form of the residual noise for the $\alpha$ variable, given in Eq.~\ref{alpha},
led us to wonder whether differences of two $\alpha$-like combinations
might be able to cancel out the laser noise that remained.  We have thus
defined what we have called $\Delta$-{\it Sagnac} variables, formed by
using each of the six one-way links twice, for a total of twelve terms.
When the general form for such a variable is expanded using Eq.~\ref{phase} for the
laser phase noise, we find that there are 13824 ($=24\times24\times24$)
possible sets of 12 equations for the 11 time delays. There are a total of 60
unique, nontrivial solutions to these sets of equations. An example of one of
these $\Delta$-Sagnac variables is
\begin{equation}\label{dalpha}
\fl \eqalign{\Delta\alpha(t)&=y_{12}(t-L_{31}-L_{23})+y_{23}(t-L_{31})+y_{31}(t)\\
&-y_{13}(t-L_{21}-L_{32})-y_{32}(t-L_{21})-y_{21}(t)\\
&-y_{12}(t-L_{21}-L_{32}-L_{13}-L_{31}-L_{23})
-y_{23}(t-L_{21}-L_{32}-L_{13}-L_{31})\\
&-y_{31}(t-L_{21}-L_{32}-L_{31})+
y_{13}(t-L_{31}-L_{23}-L_{12}-L_{21}-L_{32})\\
&+y_{32}(t-L_{31}-L_{23}-L_{12}-L_{21})+y_{21}(t-L_{31}-L_{23}-L_{12}).}
\end{equation}
Most importantly for the goal of isolating instrument noise in the
detectors, we have found one $\Delta$-Sagnac variable that recovers the
insensitivity to low-frequency gravitational waves that characterized
$\zeta(t)$.  This is
\begin{eqnarray}\label{dzeta}
\fl && \Delta\zeta(t)=y_{13}(t-L_{21}-L_{12}-L_{32})
+y_{32}(t-L_{12}-L_{23}-L_{31})+y_{21}(t-L_{23}-L_{12}-L_{32})\nonumber \\
\fl &&-y_{12}(t-L_{32}-L_{13}-L_{21})-y_{23}(t-L_{21}-L_{12}-L_{32})
-y_{31}(t-L_{32}-L_{23}-L_{12})\nonumber \\
\fl &&-y_{13}(t-L_{31}-L_{12})-y_{32}(t-L_{13}-L_{31})-y_{21}(t-L_{32}-L_{13})\\
\fl &&+y_{12}(t-L_{31}-L_{13})+y_{23}(t-L_{31}-L_{12})+y_{31}(t-L_{13}-L_{32})\nonumber
\end{eqnarray}
The $\zeta$-like insensitivity to gravitational waves of this variable may
be seen by the fact that, in the limit $L_{ij}=L_{ji}$, it may be written
\begin{equation}\label{dzetasym}
\Delta\zeta(t)=\zeta(t-L_{12}-L_{23})-\zeta(t-L_{13}), 
\end{equation}
so that $\Delta\zeta(t)$ only differs from this combination of $\zeta(t)$
terms by the small amount arising from the asymmetry in armlengths, or by
only a few parts in $10^{-6}$ of the gravitational wave signals in
each $y_{ij}$.  The subtraction in Eq.~\ref{dzetasym} will affect both the signal and
the noise in the $\zeta$ terms, so the signal-to-noise ratio for
$\Delta\zeta(t)$ will be the same as that for $\zeta(t)$.  Thus, we are
able to recover a diagnostic TDI variable, at the cost of requiring
slightly more telemetered data from each spacecraft to allow the
complicated twelve-term variable to be constructed.

\subsection{Interferometer-type Variables}

In contrast to the cases of the Sagnac and symmetric Sagnac variables, the
interferometer variables $X(t)$, $Y(t)$, and $Z(t)$, will provide output
for LISA in which laser phase noise is exactly eliminated, even when the
detector arms are rotating, though we will need to differentiate between
$L_{ij}$ and $L_{ji}$ in the proper places.  The solution displayed in
Eq.~\ref{Xsol} is valid even when $L_{ij}\neq L_{ji}$, yielding a form for $X(t)$
given by
\begin{equation}\label{Xx}
\fl \eqalign{ X(t)&= y_{12}(t-L_{31}-L_{13}-L_{21})-y_{13}(t-L_{21}-L_{12}-L_{31})
  +y_{21}(t-L_{31}-L_{13})\\
  &-y_{31}(t-L_{21}-L_{12})+y_{13}(t-L_{31})-y_{12}(t-L_{21})+y_{31}(t)-y_{21}(t)}
\end{equation}
Essentially, this form for $X(t)$ is found from the old form in Armstrong,
Estabrook, and Tinto \cite{aet} by substituting $2L_{ij}\rightarrow
L_{ij}+L_{ji}$. If we begin with Eq.~\ref{Xx} and permute indices according to
$1\rightarrow2\rightarrow3$, we find valid redefinitions for the other
interferometer variables $Y(t)$ and $Z(t)$ as well.

Using procedures like that leading to Eq.~\ref{Xx}, we have also found
redefinitions of the beacon, monitor, and relay variables that will
succeed in exactly canceling laser phase noise, even if the detector is
rotating.

\section{The Effects of Flexing}

The direction symmetry that is broken by rotation is a discrete symmetry,
but pure rigid rotation still preserves the important continuous symmetry,
$L_{ij}(t+\tau)=L_{ij}(t)$ for any value of $\tau$.  When this symmetry is
preserved, then the eight equations for seven unknowns that must be
satisfied to eliminate phase noise in Eq.~\ref{X} will contain only the four
constants, $L_{12}$, $L_{21}$, $L_{13}$, and $L_{31}$.  These few
constants create a subtle symmetry in the equations that permits the $X(t)$
solution to exist.  Unfortunately, the LISA orbits do not allow for
time-translation symmetry.  The non-circularity of the two-body orbits and
the perturbations of the orbits by the planets produce a complicated
flexing of the arms in the detector.  As a result, there is a time
dependence in the armlengths, and Eq.~\ref{phase} for the laser phase noise in a
one-way link should be written
\begin{equation}\label{phase2}
y_{ij}=\phi_i(t-L_{ij}(t))-\phi_j(t) .
\end{equation}
To see how this asymmetry complicates the equations for laser phase noise
cancellation, let us rewrite Eq.~\ref{Xe}, taking the time dependence in the
$L_{ij}$ explicitly into account
\begin{eqnarray}\label{Xtime}
X(t)&=&\phi_1(t-\lambda_1-L_{12}{}^1)-\phi_2(t-\lambda_1)\nonumber \\
&& -\phi_1(t-\lambda_2-L_{13}{}^2)+\phi_3(t-\lambda_2)\nonumber \\
&& +\phi_2(t-\lambda_3-L_{21}{}^3)-\phi_1(t-\lambda_3)\nonumber \\
&& -\phi_3(t-\lambda_4-L_{31}{}^4)+\phi_1(t-\lambda_4)\nonumber \\
&& +\phi_1(t-\lambda_5-L_{13}{}^5)-\phi_3(t-\lambda_5)\\
&& -\phi_1(t-\lambda_6-L_{12}{}^6)+\phi_2(t-\lambda_6)\nonumber \\
&& +\phi_3(t-\lambda_7-L_{31}{}^7)-\phi_1(t-\lambda_7)\nonumber \\
&& -\phi_2(t-\lambda_8-L_{21}{}^8)+\phi_1(t-\lambda_8)  ,\nonumber
\end{eqnarray}
where the notation $L_{ij}{}^k$ denotes $L_{ij}(t-\lambda_k)$.  There are eight
different constants in Eq.~\ref{Xtime} and the orbits are complicated enough that
there appears to be no orbital symmetry relating them to each other. As a
result, there are no solutions to any of the 96 possible sets of equations
that must be satisfied in order to cancel the $\phi_i$ terms identically.

To evaluate how much noise may be canceled and how much noise remains,
let us cancel as many terms in Eq.~\ref{Xtime} as we can.  We start by setting
$\lambda_7=\lambda_8=0$ and proceed from bottom to top in Eq.~\ref{Xtime}. We end up with a
definition of $X(t)$ given by
\begin{equation}\label{Xfinal}
\fl \eqalign{X(t)&=y_{12}(t-L_{31}-L_{13}^{(1)}-L_{21}^{(2)})-
  y_{13}(t-L_{21}-L_{12}^{(1)}-L_{31}^{(2)})
  +y_{21}(t-L_{31}-L_{13}^{(1)})\\
  &-y_{31}(t-L_{21}-L_{12}^{(1)})+y_{13}(t-L_{31})-
  y_{12}(t-L_{21})+y_{31}(t)-y_{21}(t),}
\end{equation}
where $L_{21}=L_{21}(t)$, $L_{31}=L_{31}(t)$,
$L_{12}^{(1)}=L_{12}(t-L_{31})$, $L_{13}^{(1)}=L_{13}(t-L_{31})$,
$L_{21}^{(2)}=L_{21}(t-L_{31}-L_{13}^{(1)})$, and
$L_{31}^{(2)}=L_{31}(t-L_{21}-L_{12}^{(1)})$.  With this definition, all
but two terms in Eq.~\ref{Xtime} cancel, leaving
\begin{equation}\label{Xres}
\fl X(t)=\phi_1(t-L_{31}-L_{13}^{(1)}-L_{21}^{(2)}-L_{12}^{(3)})
-\phi_1(t-L_{21}-L_{12}^{(1)}-L_{31}^{(2)}-L_{13}^{(3)}),
\end{equation}
where $L_{12}^{(3)}=L_{12}(t-L_{31}-L_{13}^{(1)}-L_{21}^{(2)})$ and
$L_{13}^{(3)}=L_{13}(t-L_{21}-L_{12}^{(1)}-L_{31}^{(2)})$.  To evaluate
the size of the noise that remains in Eq.~\ref{Xres}, let us estimate the time
dependence in the armlengths using $L_{ij}^{(n)}=L_{ij}(t)-nV_{ij}L$,
where $V_{ij}\approx V_{ji}$ is the rate of change of the armlength in
seconds per second and $L$ is a typical one-way light time.  Expanding
each $\phi_1$ term in Eq.~\ref{Xres} in a Taylor series, we find
\begin{eqnarray}
X(t)&=&\phi_1(t-L_{31}-L_{13}-L_{21}-L_{12})+(V_{13}L+5V_{12}L)\dot\phi_1 \nonumber \\
&&-\phi_1(t-L_{21}-L_{12}-L_{31}-L_{13})-(V_{12}L+5V_{13}L)\dot\phi_1 . 
\end{eqnarray}
Cancelling and combining, we arrive at
\begin{equation} 
X(t)=4L(V_{12}-V_{13})\dot\phi_1 , 
\end{equation}
giving a requirement on laser frequency noise of
\begin{equation}\label{Xreqi}
\sqrt{S_\nu}={\sqrt{S_X}\over4L(V_{12}-V_{13})}=
5{\rm Hz}/\sqrt{{\rm Hz}},
\end{equation}
where, in the last step, we have used $V_{12}-V_{13}=(10 {\rm m/s})/c$ and
$L=16.7$\ s.  The current science requirement for LISA specifies a laser
frequency stability of $30{\rm Hz}/\sqrt{{\rm Hz}}$.  We conclude that this
requirement needs to be strengthened by at least a factor of 6 and
probably 10.

Similar results apply to all TDI variables --- those that survive the
breaking of direction symmetry will be useful, even though they do not
survive the breaking of time-translation symmetry, but only if the laser
frequency stability requirement in Eq.~\ref{Xreqi} is satisfied.

\section{Summary}

A general property of all TDI variables is that the conditions governing
the canceling of laser phase noise involve $n$ equations for the $n-1$
time delays. In order for a solution to exist, the interferometer geometry
must posses certain symmetries. When the discrete symmetry $L_{ij}=L_{ji}$
is broken by rotation, the interferometer variables survive, but the
Sagnac variables are lost. However, the loss of a discrete symmetry can be
overcome by defining more complicated TDI combinations. On the other hand,
when the continuous symmetry $L_{ij}(t) =L_{ij}(t+\tau)$ is broken, as is
the case for the LISA orbits, {\it there are no TDI variables that
completely cancel laser phase noise}.

Thus, the first conclusion of this paper is that, without a breakthrough
in laser stability that would allow the requirement in Eq.~\ref{sagreq} to be
realized with some margin, the Sagnac variables, $\alpha(t)$, $\beta(t)$,
and $\gamma(t)$, and the symmetric Sagnac variable $\zeta(t)$, will be
dominated by laser phase noise and will not be of use in LISA data
analysis.  On the other hand, in the limit of rigid rotation, the
$\Delta$-Sagnac variables, especially the $\Delta\zeta(t)$ variable that
allows the detector to go ``off-source'', are still available, as are the
simple interferometer variables, $X(t)$, $Y(t)$, and $Z(t)$.

Our second conclusion is that laser phase noise will dominate the LISA
sensitivity window for all variables, because of the breaking of
time-translation symmetry, unless a way is found to reduce laser frequency
stability to that calculated in Eq.~\ref{Xreqi}. In this case the variables that survive
the direction asymmetry will all be useful for data analysis.
If the laser stability
requirement proves too difficult, an alternative would be to reduce the
LISA armlengths, simultaneously reducing $L$ and $V_{ij}$. This would ease
the laser stability requirement, though the resulting reduction in the
response of the detector to gravitational waves is likely to reduce the
overall sensitivity of the detector, due to other noise sources.

\ack
We would like to thank Daniel Shaddock for suggesting the rotation problem
and for subsequent discussions and Joseph Plowman and Louis Rubbo for work
defining the proper forms of the interferometer variables for
direction-asymmetric armlengths. This work was supported by the NASA
EPSCoR program through Cooperative Agreement NCC5-579.


\Bibliography{99}
\bibitem{ghtf} G. Giampieri, R.W. Hellings, M.
Tinto, and J.E. Faller, Opt.\ Commun.\ {\bf 123} 669 (1996); M. Tinto and
J.W. Armstrong, Phys.\ Rev.\ D {\bf 59} 102003 (1999).

\bibitem{ds} Daniel Shaddock, private communication.

\bibitem{ron} R.W.\ Hellings, Phys.\ Rev.\ D {\bf
64} 022002 (2001).

\bibitem{aet} J.W. Armstrong, F.B. Estabrook, and
M. Tinto, Ap.\ J.\ {\bf 527} 814-826 (1999).

\bibitem{eta} F.B. Estabrook, M. Tinto, and J.W.
Armstrong, Phys.\ Rev.\ D{\bf 62} 042002 (2002).

\bibitem{dr} K. Danzmann and A. Rudiger, Class. Quantum Grav. {\bf 20} S1-S9 (2003).

\endbib
\end{document}